\documentstyle[12pt]{article} \topmargin=-0.4in

\begin{document}
\begin{center} {\Huge \bf  New vortex solution in $SU(3)$ gauge-Higgs
theory}\\
\vskip0.2in

F.A.~Schaposnik\\ {\it Departamento de Fisica, Universidad
Nacional de La Plata, C.C. 67, (1900) La Plata, Argentina}\\ and
P.~Suranyi\\ {\it University of Cincinnati, Cincinnati, Ohio 45221-0011,
USA}\\ (Received 11 May 2000)
 \end{center}

\abstract{ Following a brief review of known vortex solutions
in $SU(N)$ gauge-adjoint Higgs theories we show the existence of a new
``minimal'' vortex solution in $SU(3)$ gauge theory with two adjoint
Higgs bosons.  At a critical coupling the vortex
decouples into two abelian vortices, satisfying Bogomol'nyi type, first
order, field equations. The exact value of the vortex energy (per unit
length) is found in terms of the topological charge that equals to the
$N=2$ supersymmetric charge, at the critical coupling. The critical
coupling signals the increase of the underlying supersymmetry. }

\section{Introduction}

Classical solutions of nonabelian gauge theories have played an important role
in a variety of contexts.~\cite{review1}  Classical solutions in Higgs theories
may play an important role in cosmology.~\cite{review2}
They may also be relevant in models of
confinement.~\cite{suranyi}  Different classical
objects may affect cosmology, symmetry breaking, etc. in different ways.
Therefore, it is of considerable importance to find
all classical solutions and investigate their properties.

Vortex solutions are solitons in 2+1 dimensions and are stringlike extended
objects in 3+1 dimenions.  In 3+1 dimensions they have infinite energy (the
energy per unit length is finite) but condensed vortices contribute a finite
amount to the free energy per unit volume. Nonabelian vortex configurations
were discussed
 in \cite{dV}-\cite{sc}; explict vortex solutions were first found
in ref.~\cite{dVS1}. The existence of nonabelian
vortices is the consequence of nontrivial topological classes in the mapping
$S_1\to SU(N)/Z_N.$ The homotopy group of this mapping is $Z_N$, implying the
existence of $N-1$ distinct stable vortices. As the symmetry, classifying
vortices, is the center of the gauge group $SU(N)$, one needs to introduce
Higgs fields that break $SU(N)$ symmetry, but not the center $Z_N$. The
smallest representation for the Higgs fields, such that they commute with
the center, is the adjoint representation.
Therefore, one needs to use one or more adjoint Higgs bosons to break the
symmetry.  Symmetry breaking induced by a single adjoint Higgs boson is not
complete. The adjoint Higgs, when diagonalized, commutes with the `diagonal'
generators, the elements of the  Cartan subgroup, $[U(1)]^{N-1}.$ The relevant
classical objects in such a theory are 't Hooft-Polyakov
monopoles.~\cite{thooft1}~\cite{polyakov} Thus, at least two adjoint Higgs
bosons
are needed to break the symmetry down to its center.

Vortex solutions found in~\cite{dVS1} correspond to $SU(N)$ adjoint
Higgs theories with $N$ Higgs bosons. In fact, one would think that a
`minimal' solution could be found with only two Higgs bosons.  The first
Higgs boson breaks the symmetry down to the maximal abelian subgroup and
then another Higgs, that is kept non-parallel with the first one, can
break all the remaining continuous symmetries.  The purpose of this paper
is to show that vortex solutions in $SU(3)$ gauge theory with two
adjoint Higgs bosons exist and to study the properties of these solutions.

The equations of motion in Abelian~\cite{bogomolny}-\cite{dVS2}
and nonabelian~\cite{CLS} vortex
model were shown to reduce to linear, Bogomol'nyi equations at
critical values of the coupling constant. This phenomenon was shown to be
related to the increase of an underlying supersymmetry of the
model.~\cite{LLW}-\cite{ENS}
The equations of motions we obtain for the $SU(3)$ Higgs theory also linearize
and decouple at critical couplings.  The relationship with increased
supersymmetry can also be ascertained as the fields decouple into a couple of
abelian vortices at the critical coupling.

In the next section we will briefly review the  solutions of field
equations for $SU(3)$ theory offered in Ref.~\cite{dVS1}. In section 3 we will
present our two Higgs model and the ansatz for solving the equations of
motion.  In section 4 we will discuss the critical coupling, the Bogomol'nyi
eqautions and their relationship to supersymmetry, followed by a concluding
section.

\section{Vortex solutions in  $SU(N)$ gauge theory with $N$ Higgs}

As usual in discussing time-independent classical solutions we will consider
the Hamiltonian, the negative of the Lagrangian in the absence of time
derivative.  The Hamiltonian   for a cylindrically
symmetric solution is of the form
\begin{equation}
H =\int
d^2x\left[\frac{1}{4}G_{\mu\nu}^2+\frac{1}{2}\sum_{A=1}^{N}(D_\mu\Phi^{(A)})^2
+ V(\Phi^{(A)})\right],
\label{ham1}
\end{equation}
Here
\begin{eqnarray}
A_\mu &=& A_\mu^a t^a \, , \;\;\;\; a=1,2,\ldots, N^2-1\nonumber\\
D_\mu &=& \partial_\mu + ie[A_\mu,~]\nonumber\\
G_{\mu\nu} &=& \partial_\mu A_\nu - \partial_\nu A_\mu + ie[A_\mu,A_\nu]
\end{eqnarray}
where $t^a$ are the $SU(N)$ generators.  We are considering
$N$ Higgs scalars $\Phi^{(A)}$ in the adjoint representation and
the potential $V[\Phi]$  chosen so
as to ensure complete symmetry breaking.

Vortex solutions to the equations of motion associated
with Hamiltonian (\ref{ham1}) have
been found in \cite{dVS1} by making an  ansatz
 that ensures non-trivial topology and
maximum symmetry breaking. Since the scalars are in the adjoint
representation, the center $Z_N$ of $SU(N)$ is the
surviving symmetry subgroup. Then,
the relevant homotopy group
for classifying topologically inequivalent
 configurations
  is non-trivial, $\pi_1(SU(N)/Z_N) = Z_N$. One then has
$N-1$ topologically non-trivial inequivalent possible solutions
which can be associated
with gauge group elements $U_n$ ($n=1,2, \ldots,N-1$ labeling the
homotopy classes). If we call $\phi$ the azimuthal angle in a plane
perpendicular  to the vortex, then $U_n(\phi)$ should
satisfy, when a turn around a closed contour is made,
\begin{equation}
U_n(\phi + 2\pi) = \exp\left( i \frac{2\pi (n + N k)}{N}\right) U_n(\phi)
\, , \;\;\;
\;\;\;
n=1,2,\ldots, N-1 \, , \;\;\; k \in Z
\label{top}
\end{equation}
Condition (\ref{top}) can be realized by writing
\begin{equation}
U_n(\phi) = {\rm diag}\left(
\exp(i(n + N k)\frac{\phi}{N}),\ldots,
\exp(i(n + N k)\frac{\phi}{N}),\exp(-i\phi\frac{N-1}{N}(n + N k)
)
\right)
\end{equation}
Then, one can construct a gauge field configuration
$A_\mu^n$ belonging to the $n$ class
 so that it satisfies, at infinity,
\begin{equation}
\lim_{\rho \to \infty} A_\mu^n = -\frac{i}{e} U_n^{\dagger}(\phi) \partial_\phi
U_n(\phi)\partial_\mu\phi = \frac{1}{e} M_n\partial_\mu \phi
\label{Mn}
\end{equation}
One has explicitly
\begin{equation}
M_n = (n + N k)\, {\rm diag} \left(\frac{1}{N},\frac{1}{N},\ldots,\frac{1}{N},
\frac{1-N}{N}
\right)
\end{equation}
and hence  $M_n$ can be written in terms of the $(N-1)$  $SU(N)$ generators
$H_i$
spanning
the Cartan subalgebra of $SU(N)$,
\begin{equation}
M_n = (n + N k) \sum_{i=1}^{N-1}m^i H_i
\label{mw}
\end{equation}
where $m^i$ are the magnetic weights, as defined in \cite{GNO}.

In view of   (\ref{Mn}), the natural ansatz for a vortex solution with
topological charge $n$ is
\begin{equation}
A_\mu^n = \frac{1}{e} \partial_\mu\phi \,M_n \, a(\rho)
\label{An}
\end{equation}
with  $a(\rho)$ such that $G_{\mu\nu} \to 0$  as $\rho \to  \infty$,
fast enough to ensure the finiteness of the energy.

The finiteness of energy also requires that, at infinity, the Higgs
scalars
$\Phi^{(A)}, (A=1,2,
\ldots,N)$    take
their vacuum value, minimizing the symmetry breaking potential. Moreover,
\begin{equation}
\lim_{\rho \to \infty} D_\mu[\Phi^{(A)}] = 0
\label{cond}
\end{equation}
Condition (\ref{cond}) can be achieved either by taking the scalars
in the Cartan algebra of $SU(N)$ or in its complement. Let us write the
$SU(N)$ generators
in the Cartan-Weyl basis, with $H_i$ the $N-1$ generators spanning the
Cartan algebra and $E_{\pm \alpha}$ those in its complement,
\begin{eqnarray}
[H_i, E_{\pm \alpha}] &=& \pm \alpha_i E_{\pm \alpha}
\nonumber\\
~[
E_{\alpha},
E_{-\alpha}]
 &=& \sum_{i=1}^{N-1} \alpha_i H_i
\label{alg}
\end{eqnarray}
where $\alpha_i = \alpha^i$ are the roots in an orthonormal basis. Then, one
can choose the symmetry breaking potential so that the first $S$  scalars
$\bar \Phi^{(1)}, \bar \Phi^{(2)}, \ldots \bar\Phi^{(S)}$ are in the
Cartan algebra and the
rest, $\Phi^{(1)}, \Phi^{(2)}, \ldots \Phi^{(T)}$
in its complement, $S+T = N$. Now,
in order to satisfy (\ref{cond}), one necessarily has
\begin{eqnarray}
\lim_{\rho \to \infty} \bar \Phi^{(A)}(\rho,\phi) &=& \sum_{j=1}^{N-1}
C_j^{(A)} H_j\nonumber\\
\lim_{\rho \to \infty}  \Phi^{(A)}(\rho,\phi) &=& U_n^{\dagger}(\phi)
\left( \sum_{\pm \alpha}
\eta^{(A)}_{\alpha} E_\alpha \right)U_n(\phi)= U_n^{\dagger}(\phi)
\eta^{(A)}U_n(\phi)
\end{eqnarray}
with $ C_j^{(A)}$ and $ \eta^{(A)}_{\alpha} $ constants. The constants
$\eta^{(A)}$ should be adjusted to so that they would mimimize
$V(\Phi^{(A)})$.

In view of the conditions described above, a consistent ansatz for a $Z_N$
vortex
configuration can be proposed in the form
\begin{eqnarray}
\bar \Phi^{(A)} &=& \sum_{j=1}^{N-1}
C_j^{(A)} H_j \nonumber\\
\Phi^{(A)} &=& U_n^{\dagger}(\phi)\left( \sum_{\pm \alpha}
\eta^{(A)}_\alpha\psi^{(A)}_\alpha(\rho) E_\alpha
\right)U_n{(\phi)}\nonumber\\ A_\phi &=& \frac{1}{e} a(\rho) M_n
\nonumber\\ A_\rho &=& A_0 = A_z = 0
\label{ansatz}
\end{eqnarray}
Here we have taken the $\bar \Phi^{(A)}$ scalars to be constant everywhere.
 $F^{(A)}_\alpha(\rho)$ and $a(\rho)$ should satisfy the
boundary conditions
\begin{eqnarray}
 \lim_{\rho \to \infty} \psi^{(A)}_\alpha(\rho) =
1\, , \;\;\;\;  & & \lim_{\rho \to \infty}a(\rho) = n.
\label{hig}
\end{eqnarray}

Ansatz (\ref{ansatz}) implies that
\begin{equation}
D_\phi\Phi^{(A)} = (n-a(\rho)) \partial_\phi \Phi^{(A)}
\label{higgseq}
\end{equation}

Given the ansatz for the $\bar \Phi$-type scalars,
the equations of motion derived from  (\ref{ham1})  take the form
\begin{eqnarray}
D_\mu G^{\mu\nu} &=& ie \sum_{A=1}^{N-1}[D_\nu
\Phi^{(A)},\Phi^{(A)}]\nonumber\\
D_\mu D^\mu \Phi^{(A)} &=& \frac{\delta V}{\delta \Phi^{(A)}}
\label{gaugeeq}
\end{eqnarray}
That is, appart from the potential, the $\bar \Phi^{(A)}$ fields play
no role in
the dynamics. Concerning the other   scalars $\Phi^{(A)}$, separability
of the equations of motion into radial and angular parts imposes \cite{dVS2}
\begin{eqnarray}
[M_n,[M_n,\Phi^{(A)}]] &=& R_n^A(\rho)\Phi^{(A)}\nonumber\\
\sum_{i=1}^{N-1}[\Phi^{(A)},[\Phi^{(A)},M_n]] &=& S_n^A(\rho) M_n
\end{eqnarray}
One can see that these conditions simplify the ansatz (\ref{ansatz}) to
\begin{eqnarray}
\bar \Phi^{(A)} &=& \sum_{j=1}^{N-1}
C_j^{(A)} H_j \nonumber\\
\Phi^{(A)} &=& \eta^{(A)}\psi^{(A)}(\rho) U_n^{\dagger}(\phi)\left(
 E_{\alpha_A}  + E_{-\alpha_A}  \right)U_n{(\phi)}\nonumber\\
A_\phi &=& \frac{1}{e} a(\rho) M_n \nonumber\\
A_\rho &=& A_0 = A_z = 0
\label{ansatz2}
\end{eqnarray}

In order to characterize the vortex solutions from the topological
point of view
one can introduce an ``electromagnetic tensor'' ${\cal G}_{\mu\nu}$
analogous to that proposed by Polyakov for the $SO(3)$ monopole
solution \cite{polyakov}. In view of the ansatz for the gauge field,
it is natural to take
\begin{equation}
{\cal G}_{\mu\nu} = \frac{{\rm tr}\left(M_nG_{\mu\nu}\right)}
{\left({\rm tr}\left(M^2_n\right)\right)^{1/2}}
\end{equation}
Then, the  flux  $\Phi$ associated to the magnetic field ${\cal G}_{12}$ reads,
for the $n$-vortex solution
\begin{equation}
\Phi = (n + N k)\Phi_0
\end{equation}
with $\Phi_0 = 2\pi/e$. Let us recall that $n=1,2, \ldots,N-1$ indicates the
topological class to which the configuration belongs while $k\in Z$
is related to
gauge transformations that, although leading to the same behavior  at infinity
(and hence are topologically trivial), cannot be well defined everywhere
and then are not gauge equivalent everywhere, thus giving, for
a fixed $n$,  different values for
the magnetic flux \cite{CLS}.

Although the analysis of the radial equations of motion and their solution can
be performed for arbitrary $N$, let us concentrate in the
$SU(3)$ vortex  solution, for which
two topologically inequivalent classes exist. The associated $U_n(\phi)$ are
(we take for simplicity $k=0$)
\begin{equation}
U_n(\phi)=e^{in\phi\lambda_8/\sqrt{3}} \, , \;\;\;\;\;  n=1,2
\label{uchoice}
\end{equation}
One then has
\begin{equation}
M_n =  \frac{n\lambda_8}{\sqrt{3}}
\end{equation}
An explicit realization of the Cartan Algebra is
\begin{equation}
H_1 = \frac{\lambda_3}{2} \, , \;\;\;\; H_2 = \frac{\lambda_8}{2}
\end{equation}
where $\lambda_3$ and $\lambda_8$ are the diagonal Gell-Mann matrices.
One then has, for the two-component magnetic weight (\ref{mw})
\begin{equation}
\vec m = (0,2/\sqrt 3)
\end{equation}
Concerning the step generators $E_\alpha$, they can be written
in terms of the Gell-Mann matrices $\lambda_i$ in the form
\begin{eqnarray}
E_{\alpha_1} + E_{-\alpha_1} &=& \frac{1}{\sqrt 2} \lambda_4 \nonumber\\
E_{\alpha_2} + E_{-\alpha_2} &=& \frac{1}{\sqrt 2} \lambda_6 \nonumber\\
E_{\alpha_3} + E_{-\alpha_3} &=& \frac{1}{\sqrt 2} \lambda_1
\label{3s}
\end{eqnarray}

The solution found in \cite{dVS1} corresponds to
 just one $\bar \Phi$-type scalar,
\begin{equation}
\bar \Phi = B \lambda_3 + C \lambda_8
\label{ty}
\end{equation}
and two $\Phi$-type ones
\begin{eqnarray}
\Phi^{(1)} &=& \frac{1}{\sqrt 6}\eta^{(1)}\psi^{(1)}(\rho)
U_n^{\dagger}(\phi)
\lambda_4  U_n(\phi)\nonumber\\
\Phi^{(2)} &=& \frac{1}{\sqrt 6}\eta^{(2)}\psi^{(2)}(\rho)
U_n^{\dagger}(\phi) \lambda_6  U_n(\phi)
\end{eqnarray}
With this choice, the radial equations of motion read
\begin{eqnarray}
\frac{1}{\rho}\frac{d}{d\rho}\left(\rho\frac{d\psi^{(A)}}{d\rho}\right)
 - \left(\frac{n - a(\rho)}{\rho}
\right)^2 \psi^{(A)} - v^{(A)}(\rho) \psi^{(A)} (\rho)  &=& 0\nonumber
\\
\rho\frac{d}{d\rho}\left(\frac{1}{\rho}\frac{da}{d\rho}\right)
- \frac{e}{2}\left(\left(\eta^{(1)}\psi^{(1)}\right)^2 +
\left(\eta^{(2)}\psi^{(2)}\right)^2\right) ({n - a(\rho)})  &=& 0
\label{syst}
\end{eqnarray}
where $v^{(A)}(\rho)$ stands for the derivative of the potential
with respect to $\Phi^{(A)}$.

The symmetry breaking potential proposed in \cite{dVS1} can be
written in the form
\begin{eqnarray}
V(\Phi^{(A)},\bar \Phi) &=& 
 \frac{g_1\eta^4}{8}\sum_{A=1}^2\left(\frac{1}{2}
{\rm Tr}[\Phi^{(A)}]^2- 1\right)^2+
\frac{\bar g_1 \eta^4}{8}\left(\frac{1}{2}{\rm
Tr}[ \bar\Phi\bar\Phi] - 1\right)^2 \label{pote}
\\
& & +\frac{g_2 \eta^4}{4}\left({\rm Tr}[\Phi^{(1)}\Phi^{(2)}]\right)^2
+ d_{abc}
\bar\Phi^a\left( \sum_{A=1}^2f_A \Phi^{(A)\,b}\Phi^{(A)\,c}
+ h \Phi^{(1)\,b}\Phi^{(2)\,c} \right)
\nonumber
\end{eqnarray}
where $\Phi^{(A)} =\Phi^{(A)\,b}\lambda^b$ and $d_{abc}$ is the completely
symmetric
$SU(3)$ tensor. In (\ref{pote}) and in our subseqent discussion we will
use Higgs fields that become normalized in the limit $\rho\to\infty$.
One can see that the choice of the same coupling constant $g_1$ for the
quartic coupling of the
$\Phi^{(A)}$ fields implies that $f^{(1)} = f^{(2)}$ and reduces
  system (\ref{syst}) to that arising in the $U(1)$ case,  which
is solved, at critical coupling,
by the solutions of the original Bogomol'nyi equations.

\section{A 2-Higgs vortex in $SU(3)$ gauge theory}

In this section we shall present a `minimal' $SU(3)$ solution with only two
Higgs fields $\Phi^{(A)}$ ($A=1,2$)  in the adjoint.
The Hamiltonian of the model is defined uniquely up to the Higgs potential.
There is a considerable freedom in the Higgs potential. In a way we
consider a Higgs potential simpler than that of the previous section, but in
an other way we generalize it such that it will disallow solutions of the
form discussed  in Ref.~\cite{dVS1}.  Vortex solutions for  a similar
generalization of the $SU(2)$ Higgs potential were shown to exist in
Ref.~\cite{suranyi}

The Higgs potential we propose is identical in form to that of
Ref.~\cite{suranyi} for $SU(2)$: \begin{equation}
V(\Phi^{(1)},\Phi^{(2)}) = \frac{g_1\eta^4}{8}\sum_{A=1}^2\left(\frac{1}{2}
{\rm Tr}[\Phi^{(A)}]^2-1\right)^2+\frac{g_2\eta^4}{4}\left(\frac{1}{2}{\rm
Tr}[ \Phi^{(1)}\Phi^{(2)}]-c\right)^2.
\label{potential}
\end{equation}
The generalization compared to Ref.~\cite{dVS1} appears in the nonzero
value of the constant
$c$, that is the cosine of the `angle' between the two Higgs fields at
infinity.  The brackets of (\ref{potential}) must vanish at infinity to
keep the Hamiltonian finite. Thus, unlike in previous models the Higgs fields
are {\em required not to be orthogonal} at infinity.  Admittedly, the model
we study
here is less general in the sense that the self coupling of the two Higgs
bosons is assumed to be identical.

The field equations derived from the Lagrangian, analogous to
(\ref{gaugeeq}) and (\ref{higgseq}), are
\begin{equation}
D_\mu G_{\mu\nu}-ie\eta^2\sum_{A=1}^2[\Phi^{(A)},D_\nu\Phi^{(A)}]=0
\label{gauge-eq}
\end{equation}
and
\begin{equation}
D_\mu D_\mu\Phi^{(A)}-g_1\eta^2\Phi^{(A)}\left(\frac{1}{2}
{\rm Tr}[\Phi^{(A)}]^2-1\right)-g_2\eta^2\Phi^{(B)}\left(\frac{1}{2}{\rm
Tr}[ \Phi^{(1)}\Phi^{(2)}]-c\right)=0,
\label{higgs-eq}
\end{equation}
where $A=1,2$ and then $B=2,1$.

As in the previous section, the ansatz we use for
finding vortex solutions is based on the philosophy that
the vortex solution is associated with a singular gauge transformation that
maps circles linked with the vortex to a smooth transformation connecting
two elements of the center.  Choosing $U_n(\phi)$, as in (\ref{uchoice}),
the Higgs fields are defined as
\begin{equation}
\Phi^{(i)}(x)=U_n(\phi)\psi^{(i)}(\rho)
U^\dagger_n(\phi),
\label{higgs1}
\end{equation}
where $i=1,2$ for the two Higgs bosons.

 The ansatz for the gauge field,
\begin{equation}
A_\mu(x)=\partial_\mu\phi[a_8(\rho)\lambda_8+a_3(\rho)\lambda_3],
\label{gauge1}
\end{equation}
is diagonal in gauge space. We will later show that unlike for vortices
of the previous section the component $a_3(\rho)$ must be different from
zero, despite the fact that this component does not contribute to the
vortex at $\rho\to\infty$. The gauge field of (\ref{gauge1}) satisfies the
gauge fixing condition
$\partial_\mu A_\mu=0$. Taking the  derivative of the Higgs field generates a
vortex contribution in the $\lambda_8$ gauge direction.  The form of the gauge
field was chosen to be able to cancel this vortex at infinity in the covariant
derivative.  Without such a cancellation the term of the Hamiltonian
containing the covariant derivative of the Higgs fields would diverge.

We still
need to show that the forms chosen for the fields are consistent with field
equations (\ref{gauge-eq}) and (\ref{higgs-eq}). Before doing so we will
further restrict the form of our solution. We will assume that the Higgs fields
have only components
\begin{equation}
\psi^{(A)}=\psi_4^{(A)}\lambda_4+\psi_6^{(A)}\lambda_6,
\label{higgs2}
\end{equation}
where $\lambda_4$ and $\lambda_6$ are off diagonal Gell-Mann matrices
matrices. Two is the minimal number of components needed to satisfy the
 all the constraints on the normalization of the Higgs fields at
$\rho\to\infty$ simultaneously. The two Higgs fields, provided their
coefficients are not identical, break
$SU(3)$ symmetry completely, down to its center,
$Z_3$.

Let us now show that the gauge structure we propose is consistent with the
field equations.  First of all consider (\ref{higgs-eq}). The two equations,
for the choices
$A=1$ and 2, are consistent with the solution
$\psi_I^{(1)}=\pm
\psi_I^{(2)}$.  We will show that the choice
\begin{equation}
\psi_4^{(1)}=\psi_4^{(2)}\equiv\psi_4, \, \, \,
\psi_6^{(1)}=-\psi_6^{(2)}\equiv\psi_6
\label{two-higgs}
\end{equation}
 is also consistent with (\ref{gauge-eq}).
Under the  assumptions (\ref{higgs1})-(\ref{two-higgs})
(\ref{gauge-eq}) can be calculated as
\begin{eqnarray}
&&D_\mu
G_{\mu\nu}-ie\eta^2\sum_{A=1}^2[\Phi^{(A)},D_\nu\Phi^{(A)}]=
\partial_\mu\phi\left[\lambda_8\rho\frac{d}{d\rho}\left(\frac{1}{\rho}\,\frac{da
_8}{d\rho}\right)+
\lambda_3\rho\frac{d}{d\rho}\left(\frac{1}{\rho}\,\frac{da_3}{d\rho}\right)\right]\nonumber\\&-&2e
\,\eta^2
\partial_\mu\phi\,
[(\psi_4)^2 e\,a_+(\sqrt{3}\lambda_8+\lambda_3)+
(\psi_6)^2
e\,a_-(\sqrt{3}\lambda_8-\lambda_3)]=0,
\label{consistency}
\end{eqnarray}
where
\begin{equation}
a_\pm=\sqrt{3}a_8+\frac{n}{e}\pm a_3.
\label{defaplus}
\end{equation}
Clearly the space and isospace structures are consistent and
(\ref{consistency})
leads to two scalar equations for the two unknown functions, $a_+$ and $a_-$.
These equations are
\begin{equation}
\rho\frac{d}{d\rho}\left(\frac{1}{\rho}\,\frac{da_+}{d\rho}\right)-4e^2\eta^2[2(
\psi_4)^2
\,a_++(\psi_6)^2
\,a_-]=0,
\label{gaugeplus}
\end{equation}
and
\begin{equation}
{\rho}\frac{d}{d\rho}\left(\frac{1}{\rho}\,\frac{da_-}{d\rho}\right)-4e^2\eta^2[
(\psi_4)^2
\,a_++2(\psi_6)^2
\,a_-]=0,
\label{gaugeminus}
\end{equation}
 In a similar way, the scalar equations reduce
to two equations for the two components,
$\psi_4$ and
$\psi_6$
\begin{equation}
\frac{1}{\rho}\frac{d}{d\rho}\left(\rho\frac{d\psi_4}{d\rho}\right)-\frac{a_+^2}
{\rho^2}\psi_4
-g_1\eta^2\psi_4(\psi_4^2+\psi_6^2-1)-g_2\eta^2\psi_4(\psi_4^2-\psi_6^2-c)=0,
\label{psi4}
\end{equation}
and
\begin{equation}
\frac{1}{\rho}\frac{d}{d\rho}\left(\rho\frac{d\psi_6}{d\rho}\right)-\frac{a_-^2}
{\rho^2}\psi_6
-g_1\eta^2\psi_6(\psi_4^2+\psi_6^2-1)+g_2\eta^2\psi_6(\psi_4^2-\psi_6^2-c)=0.
\label{psi6}
\end{equation}

The boundary conditions for the four fields are the following:
\begin{eqnarray}
&&a_\pm(0)=\frac{n}{e} \nonumber\\
&&\lim_{\rho\to\infty}a_\pm(\rho)=0\nonumber \\
 &&\psi_4(0)=\psi_6(0)=0\nonumber\\
&&\lim_{\rho\to\infty}\psi_4(\rho)=\sqrt{\frac{1+c}{2}},\,\,\,
\lim_{\rho\to\infty}\psi_6(\rho)=\sqrt{\frac{1-c}{2}}.
\label{boundary}
\end{eqnarray}

Now at this point it should be obvious that $a_3=0$, equivalent to
$a_+=a_-$ is not an admissible solution. If $a_+=a_-$ then from
(\ref{gaugeplus}) and (\ref{gaugeminus}) it follows that
$\psi_4=\psi_6$.  Such a solution would not satisfy the boundary
condition (\ref{boundary}), unless $c=0$.

Note that at $c=0$ $\psi_4=\psi_6$ and $a_+=a_-$. In other words the
$a_3$ component of the gauge field vanishes. Then, after appropriate
rescaling,  the vortex defined by (\ref{gaugeplus})-(\ref{psi6}) coincides
with that defined by (\ref{syst}), provided we choose $g_1=\bar g_1$ and
$\eta^{(1)}=\eta^{(2)}.$

We have not been able to prove analytically the existence of solutions of these
equations.  In a future publication~\cite{herat} we will study the
solutions numerically.  At special values of the couplings, however, the second
order equations become first order. The system of equations also decouples and
can be rescaled to a form identical to a combination of two critical abelian
vortices.  Abelian vortices have been well studied~\cite{dVS2} and the
existense
of solutions has been shown.

The form of the solutions for two-adjoint-Higgs model is unique up to
gauge transformations. A gauge transformation can always
bring $U_n(\phi)$ to the form used above.  Then the gauge field,
commuting with $U_n(\phi)$ should only have components $a_8$ or $a_3$.
Furthermore, the combination of constraint
\[
\sum_A [\Phi^{(A)},\partial_\mu\Phi_{(A)}]=0
\]
and of the field equations for the two Higgs fields  can only be satisfied
with at most two
nonvanishing components of $\Phi^{(A)}$. Choosing these as $\Phi_4$ and
$\Phi_6$ we arrive at the choice of this section.\footnote{Components
that can be transformed into each other by a global $U(1)\times U(1)$
transformations and therefore  satisfy the same field equations are
not counted as different. For our choice of the components $\Phi_5$ and
$\Phi_7$ can be eliminated by global  $U(1)\times U(1)$ transformations. }

\section{Critical coupling}

At a critical coupling the second order differential equations for the
gauge and Higgs
field of
 abelian vortex solutions can be transformed to linear
 equations~\cite{dVS2},\cite{bogomolny}. We will show below that the solution
found in the previous section also satisfies linear equations at a critical
coupling.  Furthermore, we will also observe that the first order equations
decouple into equations coupling the gauge field $a_+$ with $\psi_4$ and the
gauge field $a_-$ with the Higgs field $\psi_6$ only.

First of all, it will be advantageus to express Hamiltonian (\ref{ham1}) in
terms of the Higgs component $a_8$, $a_3$ (or $a_+$ and $a_-$), $\psi_4$, and
$\psi_6$. One obtains
\begin{eqnarray}
H&=&2\pi\int_0^\infty \rho\,d\rho\,\left[ \frac{1}{2\rho^4}(\rho
a_8'-a_8)^2+
\frac{1}{2\rho^4}(\rho
a_3'-a_3)^2+\frac{\eta^2}{\rho^2}(\psi_4^2a_+^2+\psi_6^2a_-^2)
\right. \nonumber\\ &+&\left. \eta^2
(\psi_4'^2+\psi_6'^2)+\frac{g_1\eta^4}{4}(\psi_4^2+\psi_6^2-1)^2
+\frac{g_2\eta^4}{4}(\psi_4^2-\psi_6^2-c)^2\right] \label{ham2}
\end{eqnarray}
The variation of (\ref{ham2}) results in field equations
(\ref{gaugeplus})-(\ref{psi6}).

Inspired by Ref. ~\cite{CLS} we write  rearrange the Hamiltonian into an
alternative form
\begin{eqnarray}
H&=&2\pi\int_0^\infty
\rho\,d\rho\,\left\{
\eta^2[{\psi_4}'
+\gamma\rho
\psi_4w_+]^2+\eta^2[\psi_6'+\delta
\rho\psi_6w_-]^2
\right.\nonumber\\
&+&\frac{1}{2\rho^4}[\rho
a_8'-a_8-\alpha \rho^2\eta^2(\psi_4^2+\psi_6^2-1)]^2+
\frac{1}{2\rho^4}[\rho a_3'-a_3-\beta\rho^2
\eta^2(\psi_4^2-\psi_6^2-c)]^2\nonumber\\&+&\left.\frac{f_1\eta^4}{4}
(\psi_4^2+\psi_6^2-1)^2+\frac{f_2\eta^4}{4}
(\psi_4^2-\psi_6^2-c)^2+\frac{1}{\rho}
\frac{dX}{d\rho}\right\},
\label{linear}
\end{eqnarray}
where $\alpha$, $\beta$, $\gamma$, and $\delta$ are yet undetermined constants
and $X$ is an undetermined form. Comparing (\ref{linear}) with (\ref{ham2})
provides the following values for the constants:
\begin{equation}
\gamma=\delta=-\frac{n}{|n|}
\label{gamma}
\end{equation}
\begin{equation}
\alpha=\sqrt{3}\beta=2\sqrt{3}e\frac{n}{|n|}.
\label{alpha}
\end{equation}
Furthermore, one obtains
\begin{equation}
X=|n|(\psi_4^2+\psi_6^2),
\label{eks}
\end{equation}
\begin{equation}
f_1=g_1-24e^2,
\label{fone}
\end{equation}
and
\begin{equation}
f_2=g_2-8e^2,
\label{ftwo}
\end{equation}
Substituting these values back into the Hamiltonian we can see that the
Hamiltonian is minimized with a minimum value of $2\pi|n|$ at the critical
couplings
\begin{equation}
g_1=24e^2,
\label{fones}
\end{equation}
and
\begin{equation}
g_2=8e^2,
\label{ftwos}
\end{equation}
if the fields satisfy the following Bogomol'nyi type equations:
\begin{equation}
\psi_4'=\frac{e}{\rho}\,\psi_4a_+,
\label{bog1}
\end{equation}
\begin{equation}
\psi_6'=\frac{e}{\rho}\,\psi_6a_-,
\label{bog2}
\end{equation}
\begin{equation}
a_8'-\frac{1}{\rho}a_8=\frac{\eta^2\rho}{2\sqrt{3}e}(\psi_4^2+\psi_6^2-1),
\label{bog3}
\end{equation}
and
\begin{equation}
a_3'-\frac{1}{\rho}a_3=\frac{\eta^2\rho}{2e}(\psi_4^2-\psi_6^2-c).
\label{bog4}
\end{equation}

In fact, appropriate linear combinations of (\ref{bog3}) and (\ref{bog4})
can be
taken to arrive at the equations
\begin{equation}
a_+'-\frac{1}{\rho}a_+=\frac{\eta^2\rho}{e}\left(\psi_4^2-\frac{1+c}{2}\right),
\label{bog5}
\end{equation}
and
\begin{equation}
a_-'-\frac{1}{\rho}a_-=\frac{\eta^2\rho}{e}\left(\psi_6^2-\frac{1-c}{2}\right),
\label{bog6}
\end{equation}

Now  field equation (\ref{bog1}) and (\ref{bog5}) decouple from (\ref{bog2})
and (\ref{bog6}). Both of these systems are identical to the systems one
obtains for the gauge and Higgs fields at critical coupling for $SU(2)$
vortices. This is not an accident, as a gauge rotation can transform $\psi_4$
(or $\psi_6$) into $\psi_1$ and $a_+$ (or $a_-$) into $a_3$ simultaneously.
In the next section we will connect the $SU(2)$ vortices with
abelian vortices at critical coupling.

\section{ Relation to Abelian vortices and supersymmetry }

One can rewrite eqs.(\ref{bog1}) and (\ref{bog5}) so
that they coincide with the Bogomol'nyi equations of an Abelian Higgs
model where $\psi_4$ is identified with the modulus of
the Higgs scalar and $a_+$ identified  with the $A_\phi$
component of the
Abelian gauge field. Indeed, if one calls
\begin{eqnarray}
f &=&  \frac{\sqrt 2 F}{e} \psi_4 \nonumber\\
  a_+- \frac{n}{e}&\equiv &  \sqrt{3}a_8+a_3= A
 \label{a}
\end{eqnarray}
then, eqs.(\ref{bog1}) and (\ref{bog5}) become
\begin{eqnarray}
f'(\rho) &=& (n + eA)\frac{f}{\rho} \nonumber\\
\frac{1}{\rho} \frac{dA_\varphi}{d\rho} &=& \frac{e}{2}\left(f^2 -
\frac{F^2}{e^2}(1+c)\right)
\label{b}
\end{eqnarray}
These are nothing but the first order (BPS)
equations, as originally written in \cite{dVS2} (see eqs. (3.5) and
(3.6) in that paper), once an axially symmetric ansatz
is imposed in the form
\begin{eqnarray}
\Phi&=& f(\rho) \exp(-in\phi)\nonumber\\
A_\varphi &=& -\frac{1}{\rho} A(\rho)
\label{c}
\end{eqnarray}
Here we have called $\Phi$ the complex scalar field and $A_\mu$
the $U(1)$ gauge field. The condition for the Higgs field at infinity
corresponds to
\begin{equation}
\lim_{\rho \to \infty} f(\rho) = \frac{F}{e}\sqrt{1+c}
\label{inf}
\end{equation}
The solution to eqs. (\ref{b}) corresponds to a vortex
with
magnetic flux 
$\Phi = -{(2n\pi)}/{e} $. First order equations (\ref{b}) solve
the second order Euler-Lagrange equations for the Abelian Higgs model with
 symmetry breaking potential
\begin{equation}
V_{U(1)} = \frac{\lambda}{8}\left(|\phi|^2 - |\phi_0|\right)^2
\label{bdv}
\end{equation}
provided
 the $\phi^4$ coupling constant $\lambda$ is chosen as \cite{dVS2},
 \cite{bogomolny}
\begin{equation}
\lambda = e^2
\label{le}
\end{equation}

We then see that eqs.(\ref{b}) correspond to the Bogomol'nyi
equations for a vortex with topological charge $-1$. Of course,
the $+1$ topological charge equation is also obtainable just by
changing eq.({\ref{a}) to
\begin{eqnarray}
f &=&   \frac{\sqrt 2 F}{e} \psi_4 \nonumber\\
 a_+ &=& -\frac{n}{e} + A
 \label{ap}
\end{eqnarray}
The same can be done for magnetic flux $n= \pm 2$ , etc.
An analogous identification can be done concerning $\psi_6$ and
$w_-$. One gets the same equations as in (\ref{b}) with $c \to -c$.

The connection between BPS relations and supersymmetry
 has been thoroughfully
analysed for the Abelian Higgs model, including the case in which
a Chern-Simons term is added to the Maxwell term \cite{LLW}-\cite{ENS}.
The outcome is that in order to achieve the $N=2$ supersymmetric extension
of the purely bosonic model, one is forced to impose the condition
(\ref{le}), exactly as it happens when trying to find a
BPS bound proceeding \`a la Bogomol'nyi~\cite{bogomolny}. In the
supersymmetric framework, the
bound for the energy coincides with the central charge of the $N=2$ SUSY
algebra, which can be seen to coincide with the magnetic flux,
related to the  topological charge.

Once the connection between the non-Abelian $SU(3)$ model presented
in Section 3 and the $U(1)$ model
is
established, the supersymmetric analysis can be done in a very simple way.
Indeed, since the first order system (\ref{bog1})-(\ref{bog6}) decouples into
two systems, one for $(w_+,\psi_4)$ and the other for $(w_-,\psi_6)$, one
can analyse them separately. Let us consider for example the former system.
Identification (\ref{a}) implies that the $U(1)$ Lagrangian from which
eqs.(\ref{bog1})
and (\ref{bog5}) can be derived takes the form
\begin{equation}
L_{U(1)} = -\frac{1}{4} F_{\mu \nu} F^{\mu \nu} + \frac{1}{2}
 (\partial_\mu \Phi^* - ie A_\mu\Phi^*)
(\partial_\mu \Phi + ie A_\mu\Phi) -
\frac{e^2}{8}\left(|\Phi|^2 - |\Phi_0|\right)^2
\label{la}
\end{equation}
In view of the axial symmetry of the problem (no $x^3$ dependence), one
should consider $\mu=0,1,2$; that is, one is effectively
working in $2+1$
dimensional space-time.  In (\ref{la}) $A_\mu$ and $\Phi$ are connected
with $w_+$ and $\psi_4$ according to eqs.(\ref{a}),(\ref{c}).

The $N=2$ supersymmetric extension of the model defined by Lagrangian
(\ref{la}) can be written in the form
\begin{eqnarray}
{ L}_{N=2} & = &   \{ -\frac{1}{4}F_{\mu\nu}F^{\mu\nu} +
\frac{1}{2}(\partial_{\mu}M)(\partial^{\mu}M) + \frac{1}{2}
(D_{\mu}\Phi)^*(D^{\mu}\Phi) - \frac{e^2}{4}M^2\vert\Phi\vert^2 \nonumber \\
& - & \frac{e^2}{8}(\vert\Phi\vert^2 -
{\Phi_0}^2)^2
+ \frac{i}{2}\overline{\Sigma}\not\!\partial\Sigma +
\frac{i}{2}\overline{\psi}\not\!\! D\psi -
\frac{e}{2} M\overline{\psi}\psi \nonumber \\
& - & \frac{e}{2}(\overline{\psi}\Sigma\Phi + h.c.)
\label{10bis}
\end{eqnarray}
Here $M$ is a real scalar,  $\psi$ and  $\Sigma$
($\Sigma = \chi  + i \xi$) being Dirac
fermions.

Lagrangian (\ref{10bis}) is invariant under the supersymmetric transformations
\begin{eqnarray}
\hat{\delta} \Sigma & = &
- \left(\frac{1}{2}\epsilon^{\mu\nu\lambda}F_{\mu\nu}
\gamma_{\lambda} + \frac{e}{2}(\vert\Phi\vert^2 - {\Phi_0}^2) +
i\not\!\partial M\right) \eta_c
\; \; \; , \; \; \; \hat{\delta} A_{\mu} = -i\overline{\eta}_c\gamma_{\mu}
\xi \nonumber  \\
\hat{\delta} \psi & = & -i\gamma^{\mu} D_{\mu}\Phi \eta_c-
(e^2)^{1/2}M\Phi \eta_c\; \; \; , \; \; \; \hat{\delta} M =
\overline{\eta}_c\chi  \; \; \; , \; \; \;
\hat{\delta} \Phi = \overline{\eta}_c\psi
\label{9}
\end{eqnarray}
The spinor supercharges generating these transformations can be shown to be
\cite{ENS}
\begin{eqnarray}
Q & = & \frac{\sqrt 2}{e\Phi_0}\int d^2x [\left(
-\frac{1}{2}\epsilon^{\mu\nu\lambda}
F_{\mu\nu}\gamma_{\lambda} + i\not\!\partial M - \frac{e}{2}
(\vert\Phi\vert^2 - {\Phi_0}^2) \right) \gamma^0\Sigma \nonumber \\
& + & \left( i(\not\!\! D\Phi)^* - \frac{e}{2}M\Phi^* \right)
\gamma^0 \psi ]
\label{14}
\end{eqnarray}
and
\begin{eqnarray}
\overline{Q} & = & \frac{\sqrt 2}{e\Phi_0}\int d^2x [
\overline{\Sigma}\gamma^0\left(-
\frac{1}{2}\epsilon^{\mu\nu\lambda}F_{\mu\nu}\gamma_{\lambda} -
i\not\!\partial M - \frac{e}{2}(\vert\Phi\vert^2 - {\Phi_0}^2)
\right) \nonumber \\
& + & \overline{\psi}\gamma^0 \left( -i\not\!\! D\Phi -
\frac{e}{2}M\Phi \right) ]
\label{15}
\end{eqnarray}
and satisfy the $N=2$ algebra
\begin{equation}
\{Q_{\alpha},\overline{Q}^{\beta}\} = 2{(\gamma_{0})_{\alpha}}^{\beta}P^{0}
+ {\delta_{\alpha}}^{\beta} Z
\label{16}
\end{equation}
where  $\alpha, \beta = 1,2$ and
\begin{equation}
P^0 = E = \frac{1}{2 e^2{\phi_0}^2}
\int d^2x \left[ \frac{1}{2}F^2_{ij} + \vert D_i\Phi\vert^2
+ \frac{e^2}{4}(\vert\Phi\vert^2 - {\Phi_0}^2)^2 \right]
\label{17}
\end{equation}
while the central charge is given by:
\begin{equation}
Z = -\frac{1}{e^2{\phi_0}^2}\int d^2x
\left[ \frac{e}{2}\epsilon^{ij}F_{ij}(\vert\Phi\vert^2 - {\Phi_0}^2)
+ i\epsilon^{ij}(D_i\Phi)(D_j\Phi)^* \right]
\label{18}
\end{equation}
Here we have considered static configurations with $A_0 = 0$ so that
$i,j=1,2$.  Moreover, we have put $M$ and all fermions to zero  to restrict
the supersymmetric model to the original $U(1)$ model.
One can easily see that the central charge, as given by (\ref{18}), coincides
with the magnetic flux,
\begin{equation}
Z = \int \partial_i\left(\frac{1}{e}A_j + \frac{i}{e^2{\Phi_0}^2}
{\Phi}^*D_j\phi \right)\epsilon^{ij} = \frac{2\pi}{e}n
\label{18bisbis}
\end{equation}

It is now easy to find  the Bogomol'nyi bound from the
supersymmetry algebra (\ref{16}).
Indeed, since the anticommutators in
(\ref{16}) are Hermitian, one has:
\begin{equation}
\{Q_{\alpha},\overline{Q}^{\beta}\}\{Q^{\alpha},\overline{Q}_{\beta}\}
\geq 0
\label{19}
\end{equation}
or using (\ref{16}),
\begin{equation}
E \geq \vert Z \vert
\label{20}
\end{equation}

In order to explicitly obtain Bogomol'nyi equations (saturating the
energy bound) from the supersymmetry
algebra, one considers
\begin{equation}
Q_{I} = \frac{Q_+ + i Q_-}{\sqrt{2}}
\label{22}
\end{equation}
\begin{equation}
Q_{II} = \frac{\overline{Q}^+ + i\overline{Q}^-}{\sqrt{2}}
\label{23}
\end{equation}
where we have defined $Q_{\pm}$ from
\begin{equation}
Q = \left( \begin{array}{c} Q_+ \\ Q_- \end{array} \right)
\label{24}
\end{equation}
\begin{equation}
\overline{Q} = \left( \overline{Q}^+ \; \; \; \overline{Q}^- \right)
\label{25}
\end{equation}
Now, suppose that a field configuration $\vert B\rangle$  saturates the
Bogomol'nyi bound derived from (\ref{19}). Then, one necessarily has
\begin{equation}
\left( Q_{I} \pm Q_{II} \right) \vert B\rangle = 0
\label{26}
\end{equation}
or, using (\ref{22})-(\ref{25}) and (\ref{14})-(\ref{15})
\begin{eqnarray}
\epsilon^{ij}F_{ij} & = & \pm {e} (\vert\Phi\vert^2 -
{\Phi_0}^2) \nonumber \\
i\epsilon_{ij}D^i\Phi & = & \pm (D_j\Phi)^*
\label{27}
\end{eqnarray}
which are nothing but the equations (\ref{b}) once the axially
 symmetric ansatz (\ref{c}) is imposed.

\section{ Conclusions}

A new vortex solution was shown to exist in $SU(3)$ gauge theory with two
adjoint Higgs bosons. This can be contrasted with the the solution found in
Ref. ~\cite{dVS1} that requires three adjoint Higgs bosons.  At a critical
value of the Higgs self-coupling (where the gauge and Higgs masses
coincide) the
Hamiltonian has an exact lower bound and the Higgs and gauge fields satisfy
first order
Bogomol'nyi type field equations.  The field equations for two Higgs and
two gauge
compenents also decouple at the critical couplings and both of the
decoupled sets are
equivalent to an $SU(2)$ vortex model at critical copupling.~\cite{suranyi}
That model, as  it has been shown here, is equivalent to an
Abelian Higgs model at critical coupling. ~\cite{dVS2}  Thus, the critical
$SU(3)$ model is ultimately equivalent to a pair of critical Abelian Higgs
models.  This relationship connects our models to supersymmetry.  The
supersymetric version of our model implies that the vortex mass per unit length
is bounded by the $N=2$ SUSY central charge, which, at the same
time equals to the magnetic flux of the vortex. In this respect, we expect
that the non Abelian vortices discussed here could play a relevant role in the
confinement scenario arising in strongly coupled supersymmetric theories
\cite{AG}.

\vspace{0.5 cm}

\noindent{\Large \bf Acknowledments}

\vspace{0.2 cm}

F.A.S will like to thank J.~Edelstein, G.~Lozano and C.~N\'u\~nez for
discussions and helpfull comments.
F.A.S is partially supported by CICBA as Investigador, and through
grants CONICET (PIP 4330/96), and ANPCYT (PICT 97/2285).  P.S. is
supported in part by the U.S. Department of Energy through grant
\#DE FG02-84ER-40153.


\begin{thebibliography}{99}
\bibitem{review1} R.~Rajaraman, {\it Solitons and
Instantons} (Elsevier, Amsterdam, 1982); C.~Rebbi and G.~Soliani, {\it
Solitons and Particles} (World Scientific, Singapore,1984).
\bibitem{review2} See for example: A.~Vilenkin and E.P.S.~Shellard, {\it
Cosmic strings and other topological deffects} (Cambridge University
Press, Cambridge,
 1994).
\bibitem{suranyi} P. Suranyi,  Phys. Lett. to be published, hep-lat/99120.
\bibitem{dV} H.~de Vega, Phys.  Rev. {\bf D18}, 2932 (1978).
\bibitem{Ha} P.~Hasenfratz, Phys. Lett. {\bf B85}, 338 (1979).
\bibitem{sc} A.S.~Schwartz and Yu S.~Tyupkin, Phys. Lett. {\bf B90}, 135
(1980).
\bibitem{dVS1} H.J.~de Vega and F.A.~Schaposnik, Phys. Rev. Lett. {\bf
56}, 2564 (1986);  Phys.  Rev. {\bf D34}, 3206 (1986).
\bibitem{thooft1} G. 't Hooft, Nucl. Phys. {\bf B79}, 276 (1974).
\bibitem{polyakov} A. Polyakov, JETP Lett. {\bf 20}, 194 (1974).
\bibitem{bogomolny} E.B.~Bogomol'nyi, Sov. J. Nucl. Phys. {\bf 24}, 449
(1976).
\bibitem{dVS2} H.J.~de Vega and F.A.~Schaposnik, Phys.
Rev. {\bf D14}, 1100 (1976).
\bibitem{CLS} L. F.~Cugliandolo, G.~Lozano, and F. A.~Schaposnik, Phys. Rev.
{\bf D40}, 3440 (1989).
\bibitem{LLW} C.~Lee, K.~Lee and E.~Weinberg, Phys. Lett. {\bf B243},105
(1990).
\bibitem{ENS} J.~Edelstein, C.~N\'u\~nez and F.A.~Schaposnik,
Phys. Lett. {\bf B329}, 30 (1994).
\bibitem{GNO} P.~Goddard, J.~Nuyts and D.~Olive, Nucl. Phys. {\bf B125}, 1
(1977).
\bibitem{herat} A. Herat, R. Rademacher, and P. Suranyi, in preparation.
\bibitem{AG} See for example: L.~Alvarez-Gaum\'e and F.~Zamora,
{\it  Trends in Theoretical Physics}, eds. H.~Falomir et al. (American
Institute of Physics, New York, 1998) and references therein.
\end{thebibliography}
 \end{document}